\documentclass[letterpaper]{article}

\usepackage[T1]{fontenc}
\usepackage[utf8]{inputenc}

\usepackage{geometry}
\usepackage{setspace}
\usepackage[numbers,sort&compress]{natbib}

\usepackage{graphicx}
\usepackage{float}
\newfloat{scheme}{htbp}{los}
\floatname{scheme}{Scheme}
\floatname{chart}{Chart}
\newfloat{graph}{htbp}{loh}

\usepackage{chemformula} 
\usepackage[version = 4]{mhchem} 

\usepackage{authblk}

\author[1]{G. Lee}
\author[1]{A. Borel}
\author[2]{T. Taniguchi}
\author[3]{K. Watanabe}
\author[1]{F. Sirotti}
\author[1]{F. Cadiz}
\affil[1]{Laboratoire de Physique de la Matière Condensée, CNRS,
Ecole polytechnique, Institut Polytechnique de Paris, 91120 Palaiseau, France}

\affil[2]{Research Center for Materials Nanoarchitectonics, National Institute for Materials Science, 1-1 Namiki, Tsukuba 305-0044, Japan}

\affil[3]{Research Center for Electronic and Optical Materials, National Institute for Materials Science, 1-1 Namiki, Tsukuba 305-0044, Japan}

\title{High-Temperature Activation of Single-Photon Emitters in monolayer WS$_2$}

\date{*Email: fabian.cadiz@polytechnique.edu}

\begin{document}

\maketitle

\begin{abstract}
Controlled activation of defect-bound excitonic states in two-dimensional semiconductors provides a route to isolated quantum emitters and a sensitive probe of defect physics. Here we demonstrate that \textit{in situ} high-temperature annealing of hBN-encapsulated monolayer WS$_2$ on a suspended microheater leads to the emergence of spectrally isolated single-photon emitters at cryogenic temperatures. Annealing at temperatures around 1100 K produces a sharp emission line, $X_L$, red-shifted by approximately 80 meV from the neutral exciton and exhibiting a linewidth below 200 $\mu$eV. Photoluminescence excitation spectroscopy and power-dependent measurements show that $X_L$ originates from annealing-induced defects in the WS$_2$ monolayer, while second-order photon correlation measurements reveal clear antibunching with $g^{(2)}(0)<0.5$. These results establish high-temperature \textit{in situ} annealing as a controlled means to access defect-bound excitonic states and single-photon emission in van der Waals materials.

\end{abstract}




\section{Introduction}
The reduced dielectric screening and strong quantum confinement in monolayer transition metal dichalcogenides (TMDs) lead to a remarkable enhancement of the Coulomb interaction \cite{velicky-2017}, such that excitons dominate their optical properties even at room temperature \cite{ugeda-2014}. Owing to their two-dimensional (2D) nature and small radius \cite{goryca-2019}, excitons in TMDs and other 2D materials are highly sensitive to local perturbations. As a result,  defects, strain, and dielectric disorder can efficiently confine excitons \cite{liang-2021}, enabling the engineering of point-like defects for the generation of single-photton emitters in atomically-thin layers.\\

Over the past decade, intense efforts have been devoted to understanding and controlling single-photon emission in 2D semiconductors \cite{carbone-2025}. Following the first observations of single-photon emitters in monolayer WSe$_2$ \cite{tonndorf-2015,srivastava-2015}, a variety of deterministic approaches have emerged, mainly based on strain engineering and electron/ion irradiation. Focused helium ion irradiation has enabled the creation of spatially ordered arrays of single-photon emitters in  monolayer MoS$_2$ with nanometric precision \cite{hotger-2023,barthelmi-2020}, although the efficiency of this method is significantly reduced in the similar material WS$_2$ \cite{micevic-2022}. Electron-beam irradiation has also been employed to activate individual colour centers in hexagonal boron nitride with high lateral precision \cite{fournier-2021}. Alternatively, thermal annealing under controlled environments has recently been shown to generate defect-related emission in monolayer MoS$_2$ \cite{mitterreiter-2021} and ultra-sharp green emitters in hBN under oxygen atmosphere \cite{fartas-2025}.\\

 In this work, we adopt this latter approach and demonstrate the generation of single-photon emitters in hBN-encapsulated WS$_2$ through high-vacuum thermal annealing. We fabricate hBN/\ce{WS2}/hBN heterostructures on suspended SiC micro-membranes that act as local micro-heaters. Integration into a cryostat enables Joule heating of the membrane up to $\sim$1200 K while maintaining the surrounding environment at cryogenic temperature. We show that this controlled thermal treatment leads to the formation of individual quantum emitters in monolayer \ce{WS2}. We measure a  recombination time in the nanosecond range at cryogenic temperatures and perform second-order photon correlation measurements yielding $g^{(2)}(0) \approx 0.4$, confirming the single-photon nature of the emission. Photoluminescence excitation spectroscopy (PLE) further demonstrates that the emitting defects originate from the TMD layer rather than from the hBN or the SiC membrane. The observed single-photon emission is attributed to excitons localized at annealing-induced defect sites, whose microscopic origin remains to be elucidated.


\section{Methods}
Fully encapsulated \ce{WS2} monolayers were prepared by mechanical exfoliation of bulk crystals followed by the assembly of hBN/\ce{WS2}/hBN heterostructures on \ce{Si/SiO2} substrates through a dry stamp transfer technique \cite{castellanos-gomez-2014}. The stack was then picked up using a polycarbonate (PC) stamp and transferred onto 120 nm-thick suspended SiC membranes by thermal release at $\sim 200^{\circ}$C, the resulting heterostructure is schematically shown in Fig. \ref{fig:fig1}(a). Figure \ref{fig:fig1}(b) shows an optical image under white light illumination of an hBN encapsulated monolayer WS$_2$ transferred onto a SiC membrane. The latter is suspended between two silicon contacts allowing for a current to be applied accross it. \\

Unless otherwise stated, the laser source was a pulsed supercontinuum laser coupled to a  tunable filter (typical FWHM of 2 nm). The laser pulses have a temporal width of 40 ps, repetition rate of 78 MHz, and were focused onto the sample by an apochromatic objective (NA=0.82) producing a spot of  $\sim 1 \;\mu$m diameter.  The photoluminescence was collected by the same objective, dispersed by a grating spectrometer and imaged by a Peltier-cooled CCD camera. For time-resolved and photon-correlation measurements, the photoluminescence was filtered by a double spectrometer and focused into a fibered Hanbury-Brown-Twiss setup. The detectors at the outputs were two identical single-photon avalanche photodiodes of temporal resolution close to 30 ps.\\

\section{Results}

\begin{figure}[H]
  \centering
  \includegraphics[width=0.8\linewidth]{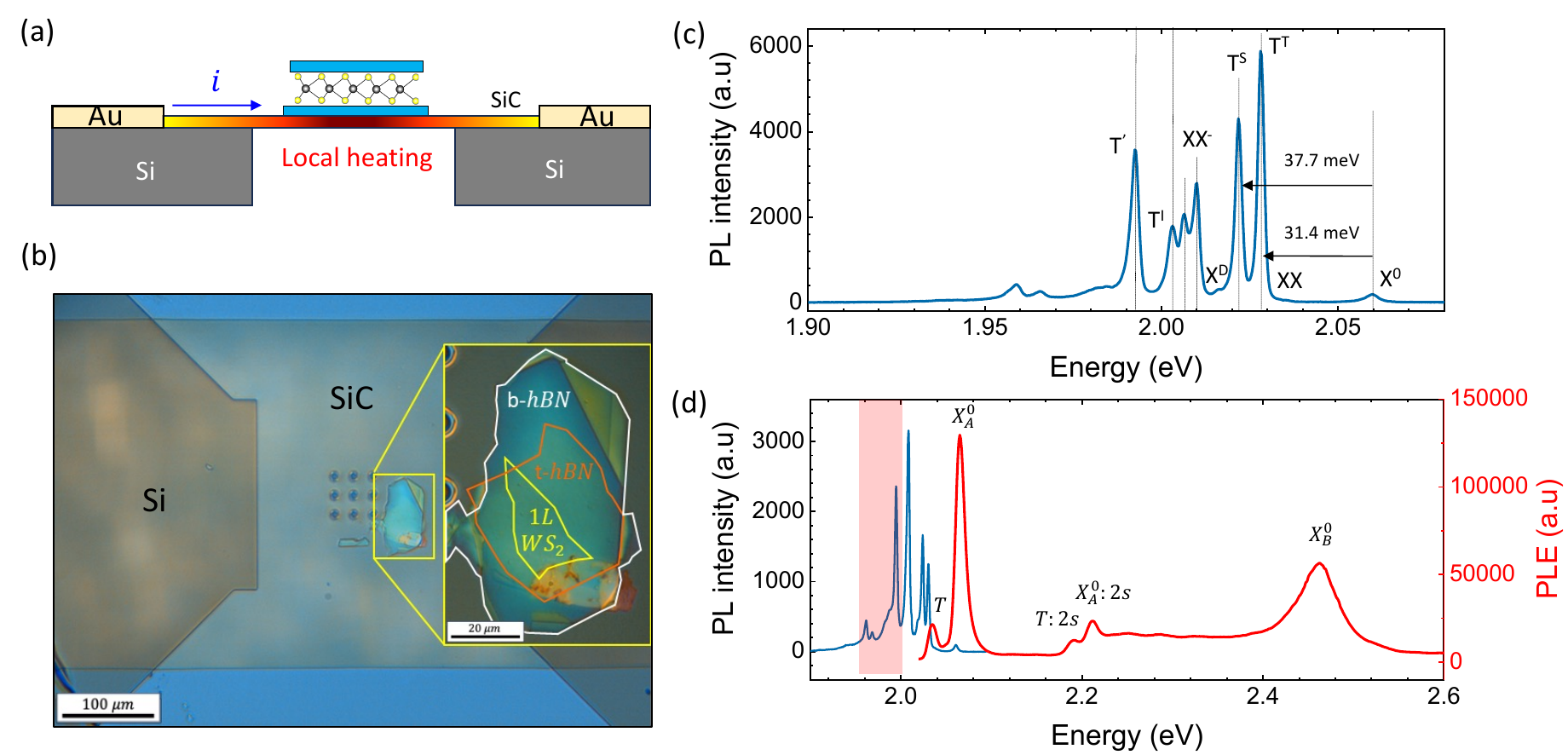}
  \caption{\textbf{Optical characterization of hBN-encapsulated monolayer \ce{WS2} on a micro-heater membrane.}
  (a) Schematic of the micro-heater chip showing the suspended SiC membrane between metallic contacts and the hBN/\ce{WS2}/hBN stack transferred on top.  
  (b) Optical image of a representative encapsulated flake on the membrane.  
  (c) Low-temperature PL spectrum at 3.6K displaying sharp neutral-exciton, trion, and dark-state related resonances with linewidths of a few meV ($\lambda_{\mathrm{laser}} = 514.5~\mathrm{nm}$ (CW), power $P_{\mathrm{laser}} = 1~\mu\mathrm{W}$).
  (d) PLE spectrum of the highlighted low-energy region exhibiting well-resolved A- and B-exciton absorption resonances and excited excitonic states ($P_{\mathrm{laser}} = 4~\mu\mathrm{W}$).}
  \label{fig:fig1}
\end{figure}

Figure \ref{fig:fig1}(c) shows the low temperature photoluminescence (PL) spectrum of one of the devices, showing sharp excitonic features with well-resolved fine structure, including the neutral exciton $X^0$, the negatively charged trions $T^T$ and $T^S$, the dark-exciton $X^D$, the negatively-charged biexciton $XX^-$, as well as several phonon assisted recombination of dark trions. The linewidths, of only a few meV, confirm the excellent optical quality of the samples and the effectiveness of encapsulation in suppressing dielectric disorder \cite{cadiz-2017}. These peaks are consistent with those previously reported in the literature  \cite{zinkiewicz-2021b, paur-2019}, and provide a direct reference for identifying defect-localized emission after exposing the sample to high temperatures.\\

Photoluminescence excitation (PLE) spectroscopy, in which the detection energy is fixed while the excitation energy is scanned at constant power, probes the absorption and relaxation pathways that populate a given emission line. A significant increase in PL intensity indicates that the excitation energy is resonant with an excitonic transition, allowing PLE to map the absorption spectrum linked to that state. Our PLE data in Fig. \ref{fig:fig1}(d), corresponding to the emission of the low-energy part of the spectrum, reveals clear resonances at the A and B excitons and at the 2s excited states of the A exciton and the trion.  These distinct optical resonances, intrinsic to pristine WS$_2$, will be important for the characterisation of single-photon emitters.

\subsection{Thermal annealing and emergence of single-photon emitters}

Thermal annealing was performed \textit{in situ} by driving a dc current through the suspended SiC membrane inside the cryostat. In order to determine the maximum annealing temperature above which all spectral signatures of WS$_2$ are destroyed, we performed Raman spectroscopy on an already calibrated micro-membrane. We found that the Raman features of monolayer WS$_2$ survive up to $\sim 1220$ K (see Fig. S1 of the Supporting Information). Since the $A_{1g}$ mode is spectrally isolated, its position can be accurately determined at each temperature. The resulting temperature dependence is shown in Fig. S2 and exhibits a clear linear behaviour up to $950^{\circ}$C, with a shift of $-0.0134\, \mbox{cm}^{-1}/ K$, providing an independent and reliable method for temperature calibration. After $1273$ K we did not detect any measurable Raman signature and the monolayer dissappeared  under white-light illumination (see Fig. S3). This is in agreement with the early work \cite{brainard-1968} of Brainard who reported the dissociation of WS$_2$ at approximately 1313 K under high vacuum. \\

\begin{figure}[H]
  \centering
  \includegraphics[width=0.8\linewidth]{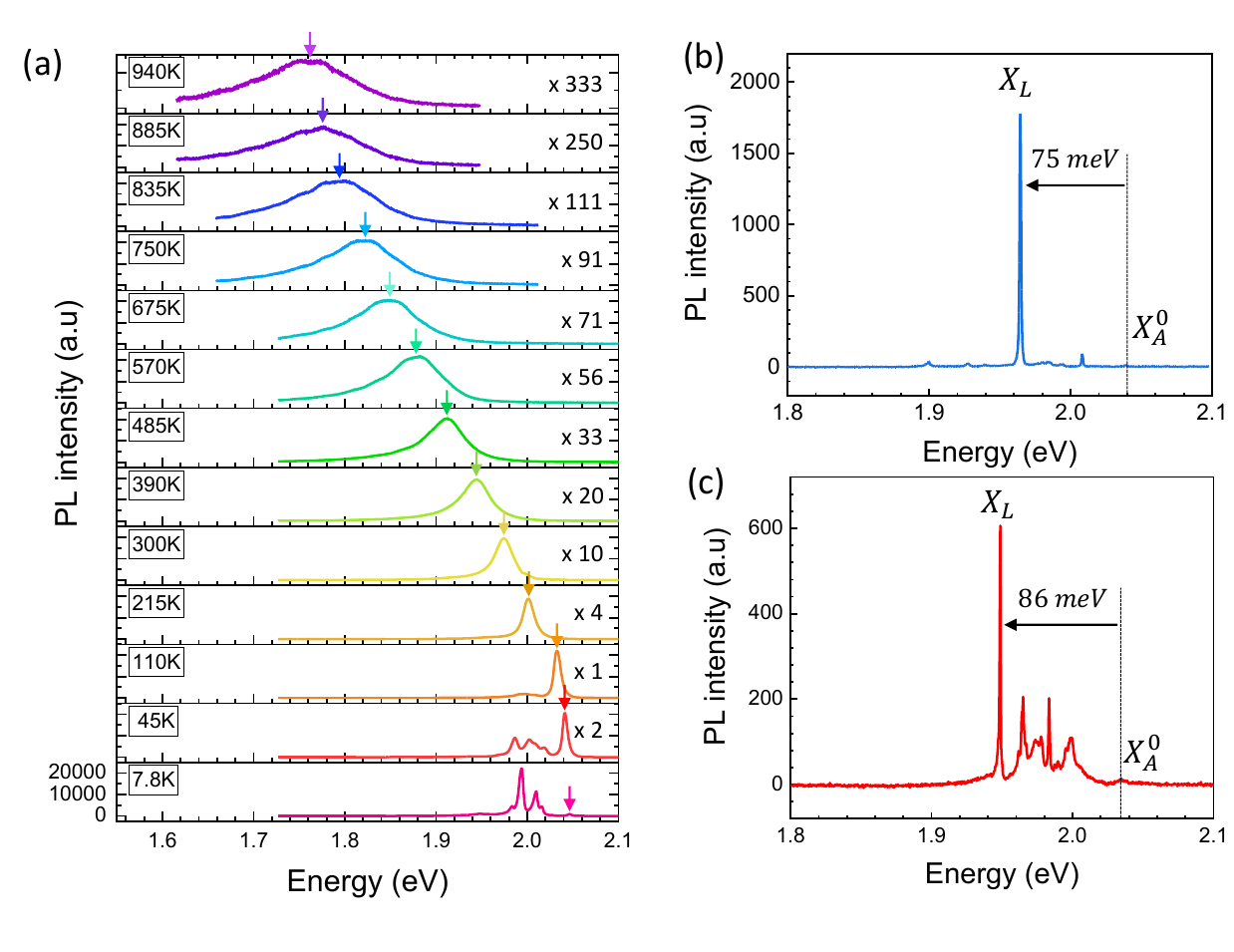}
  \caption{\textbf{Thermal-annealing-induced single-photon emission in encapsulated \ce{WS2}.}
  (a) PL spectra at different temperatures, showing the red-shift of neutral exciton energy with temperature used for temperature calibration.
  (b, c) PL spectrum after annealing around 1100~K of two samples measured ($\lambda_{\mathrm{laser}} = 514.5$ nm  (CW),\;
$P_{\mathrm{laser}} = 1~\mu\mathrm{W},\;
T = 3.65~\mathrm{K}$) in panel (b) and ($\lambda_{\mathrm{laser}} = 503~\mathrm{nm}$ (pulsed), $
P_{\mathrm{laser}} = 1~\mu\mathrm{W},\;
T = 3.72~\mathrm{K}$) in panel (c).}
  \label{fig:fig2}
\end{figure}

For each annealing temperature, the current was gradually increased in steps and held for typically 30~min at the desired value,  then turned off gradually after which the membrane cooled back to the base temperature of a few kelvin.  Importantly, we have checked that during this temperature cycle the membrane does not move significantly and the target sample is still in focus (Fig. S4). This allows us to probe the optical response of the same spot in the monolayer after several annealing steps. As mentioned earlier, the temperature of the sample can be extracted from the position of the $A_{1g}$ Raman peak. However, a good signal to noise ratio requires long acquisition times, specially at high temperatures. Instead, for the temperature calibration we have measured the PL spectrum at each annealing step as shown in Fig. \ref{fig:fig2}(a) and used Pässler's model \cite{passler-1997}:

$$E_g(T) = E_g(0)-\frac{\alpha\Theta}{2} \left[\sqrt[p]{1+\left(\frac{2T}{\Theta}\right)^p}-1\right]$$

 to extract the temperature from the peak position $E_g(T)$, indicated by the arrows in Fig. \ref{fig:fig2}(a). We have first validated, using an already temperature-calibrated chip, that Pässler's formula correctly predicts the peak position up to $\sim$900~K with the exact same parameters ( $\alpha \approx 3.47 \cdot 10^{-4} eV/K$, $\Theta \approx 200.89K$ and $p \approx 2.35$) previously reported for WS$_2$ \cite{nagler-2018a}, except for the exciton position at 0 K which is different in encapsulated monolayers due to the change of the dielectric environment ($E_g(0) \approx 2.047 eV$ for the sample shown in Fig. \ref{fig:fig2}). For temperatures higher than 950 K, the blackbody radiation of the membrane becomes more intense than the PL itself in the region in which both overlap. In that case we use Planck's law to determine the sample temperature. We have checked that at a temperature at which both the PL and the blackbody radiation are discernable, both methods yield the same temperature up to $\pm 5$ K (see the Supporting Information). \\

 After annealing up to $\sim$1000~K, the PL spectra at cryogenic temperature remain dominated by the intrinsic excitonic emission, apart from redshifts and intensity changes which could be attributed to strain induced by the annealing treatment.  
  After annealing around 1100~K, however, an ultra sharp emission appears on the low-energy side of $X^0$. Figure \ref{fig:fig2}(b,c) show the emergence of such narrow emissions in two different samples. These lines, denoted here by $X_\mathrm{L}$, were located 75 meV and 86 meV  below the neutral exciton, respectively. In contrast to the few meV linewidth of the other excitonic peaks observed in as-exfoliated samples, $X_\mathrm{L}$  exhibits a linewidth below 0.2~meV at low excitation power, limited here by our spectrometer's resolution.

\subsection{Excitonic origin and recombination dynamics of $X_\mathrm{L}$}

\begin{figure}[H]
  \centering
  \includegraphics[width=0.8\linewidth]{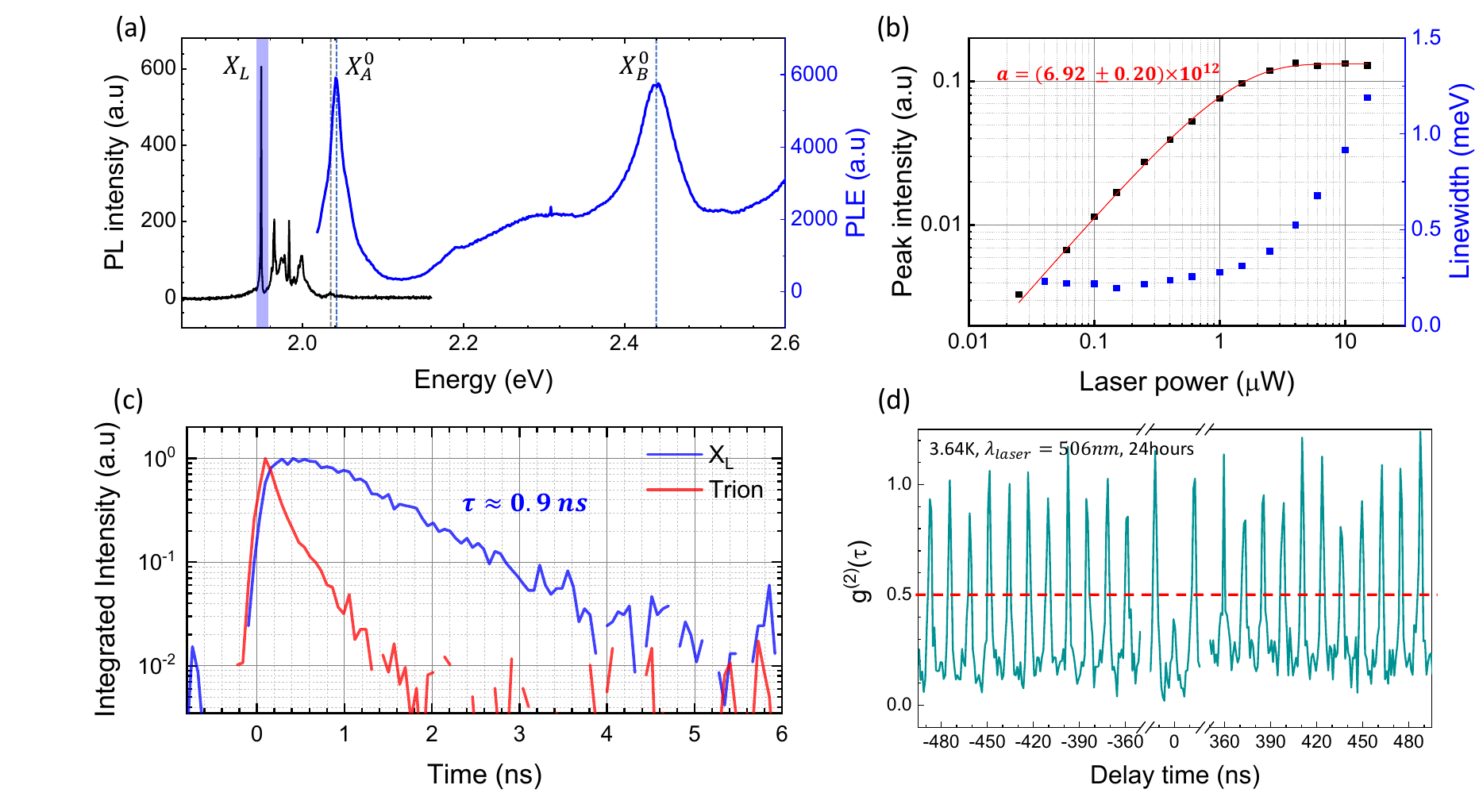}
  \caption{\textbf{Optical signatures of the SPE $X_\mathrm{L}$.}
  (a) PLE spectrum recorded at the $X_\mathrm{L}$ emission energy, showing strong resonances at the A and B exciton absorption energies.  
  (b) Power dependence of the $X_\mathrm{L}$ intensity and linewidth, exhibiting saturation of the intensity and resolution-limited linewidth at low power. ($\lambda_{\mathrm{laser}} = 503 ~\mathrm{nm}$ (pulsed)).
  (c) Time-resolved PL traces of $X_\mathrm{L}$ and trion emission, highlighting the longer lifetime of the localized state. ($\lambda_{\mathrm{laser}} = 506~\mathrm{nm}$ (pulsed), power $P_{\mathrm{laser}} = 2.5~\mu\mathrm{W}$).  
  (d) Second-order correlation function $g^{(2)}(\tau)$ under pulsed excitation, displaying pronounced antibunching with $g^{(2)}(0) < 0.5$. ($\lambda_{\mathrm{laser}} = 506~\mathrm{nm}$ (pulsed), power $P_{\mathrm{laser}} = 1~\mu\mathrm{W}$).}
  \label{fig:fig3}
\end{figure}

We have performed PLE spectroscopy on the $X_\mathrm{L}$ emission while scanning the laser energy across the excitonic resonances. In both samples, the resulting PLE spectrum (shown in Fig. \ref{fig:fig3}(a) for one of them) exhibit the two main absorption resonances intrinsic to monolayer WS$_2$: the A and B exciton transitions. This mirrors the PLE recorded in the as-exfoliated WS$_2$ (Fig. \ref{fig:fig1}(d)), and demonstrates that $X_\mathrm{L}$ emission is enhanced whenever free excitons are generated in the \ce{WS2} layer, excluding other possible origins such as termally-induced defects in the hBN ecapsulation layers or in the SiC membrane. \\

In addition, power-dependent PL measurements (shown in Fig. \ref{fig:fig3}(b)) confirm the defect nature of $X_\mathrm{L}$. Indeed, the integrated intensity increases linearly with excitation power before saturating, as expected for a small number of defect sites that can host at most one exciton at a time.
The continuous line in Fig. \ref{fig:fig3}(b) corresponds to a fit of the form:

$$
I_{\mathrm{PL}}(P_{\mathrm{avg}}) = C \left( 1 - e^{-\frac{\alpha}{f_{\mathrm{rep}}} P_{\mathrm{avg}}} \right)
$$

where $f_{\mathrm{rep}}$ denotes the laser repetition rate, $P_{\mathrm{avg}}$ represents the average laser power, $C$ is a constant, and $\alpha$ is the fitting parameter. The justification of this model is found in the Supporting Information.
 The value of $\alpha$ allows us to estimate that the probability to excite the defect is $p=2.13 \times 10^{-6}$
 per incident photon. Moreover, the number of counts at saturation allows us to estimate the quantum yield of the emitter. This is because the measured lifetime (to be discussed below) is much longer than the pulse duration, yet much shorter than the interval between pulses. At saturation, at most one photon per pulse can be therefore emitted. By taking into account the losses of the optical setup, we estimate the quantum yield to be  $\sim 5 \;\%$.
Figure \ref{fig:fig3}(b) also shows that the linewidth of $X_\mathrm{L}$ broadens  with power, possibly due to local heating caused by the laser.\\

Figure \ref{fig:fig3}(c) shows the time-resolved PL (TRPL) measurement of $X_\mathrm{L}$ emission. It decays with a characteristic lifetime of $\sim$0.9~ns. This lifetime is significantly longer than that of trion emission (typically 100 ps) in the same sample and similar to the lifetime typically reported for single-photon emitters in 2D materials \cite{srivastava-2015,carbone-2025}. Finally, the quantum nature of $X_\mathrm{L}$ is probed through second-order photon correlation measurements using a Hanbury Brown-Twiss setup under pulsed excitation. The measured intensity correlation function $g^{(2)}(\tau)$ , shown in Fig. \ref{fig:fig3}(d), shows a clear antibunching dip at zero delay with $g^{(2)}(0)\approx 0.4$ without any background correction procedure. This unambiguously proves that $X_\mathrm{L}$ acts as a single-photon emitter. The moderate residual value of $g^{(2)}(0)$ is attributed to background emission and imperfect spectral filtering, and could be further reduced by improving the collection optics (to decrease the laser excitation power) and the spectral selectivity.

\section{Discussion}
Although the generation of defects (not necessarily single-photon emitters) by thermal annealing has already been reported in other two-dimensional materials, the high temperature required here (above 1000 K) for the observation of single-photon emitters is uncommon. For example, recent results on monolayer MoS$_2$ showed that defect emission attributed to sulfur vacancies already becomes visible after annealing at 500 K, and that the photoluminescence completely vanishes above 900 K \cite{mitterreiter-2021}. If the nature of the defect observed here is also linked to chalcogen vacancies (or any complex involving such defect), our results can be explained by their significantly higher activation energy in WS$_2$ compared to MoS$_2$. This is consistent with the stronger W–S bond compared to the Mo–S bond and with calculations that put the chalcogen vacancy formation energy in WS$_2$ as the highest of the TMD family \cite{guo-2015}. Our results further show that the photoluminescence of WS$_2$ can withstand thermal treatments at substantially higher temperatures than MoS$_2$.\\

 We have also observed another defect-related emission in several of our samples after annealing at 800 K, as shown in Fig. S6 and Fig. S7 of the Supporting Information. This line, denoted here by $X'_L$, is red-shifted by 120 meV with respect to the neutral exciton and has a 3 meV linewidth at low temperatures. This is closer to the redshift typically observed in MoS$_2$ and consistent with a recent work that attributes emission in this energy range to paired-sulfur vacancies in electron-irradiated WS$_2$ \cite{sun-2024}. Although we did not observe anti-bunching in photon-correlation experiments for this line, we can't exclude that single-photon emission is possible if only a single of such defect is isolated under the laser spot. The smaller annealing temperature required to observe this line compared to the $X_L$ may indicate that $X'_L$ occur due to migration of already present sulfur vacancies, known to be present in large concentrations in as-exfoliated monolayers \cite{vancso-2016}. \\
 
 It is worth mentioning that, in non-encapsulated monolayers, only this latter type of defect line was  observed, this time red-shifted by 220 meV due to a smaller dielectric screening compared to the encapsulated samples. However, the emission was not stable in time nor under pulsed laser excitation. This highlights the importance of encapsulation for the observation and characterisation of these kind of defects.  In addition to limiting chemical contamination (and possibly oxygen passivation of vacancies), hBN encapsulation is also known to suppress photodoping effects \cite{cadiz-2016a} and charge noise, which likely contributes to the high temporal stability and narrow linewidth of the observed emitters in encapsulated monolayers. This strongly indicates that encapsulation is a necessary ingredient to stabilize the charge environment of defects and to enable single-photon emission in WS$_2$.\\
 
While Helium-ion bombardment of encapsulated monolayer WS$_2$ has been recently shown to produce narrow emission, this occurs in the near infrared \cite{micevic-2022}, $\sim 500$ meV below the exciton emission. We did not observe this kind of emission after annealing, and the large energy difference with respect to the neutral exciton may indicate a very different nature for those defects  compared to the $X_L$ and $X'_L$ discussed here.  
Note finally a crucial advantage of our \textit{in situ} annealing approach with respect to standard annealing procedures: after each annealing step, any possible adsorbates are efficiently desorbed and cryopumped by the cold surfaces of the cryostat, preventing the thermally-induced defects from interacting with these molecules.\\

 To unambiguously identify the microscopic structure — vacancy, divacancy-impurity complex or other configuration — one promising route is to perform scanning tunneling luminescence microscopy, allowing atomic-scale imaging and simultaneous optical readout \cite{huberich-2025}. Furthermore, polarization-resolved photoluminescence (PL) studies of the emitters are required to establish whether they preserve the spin–valley coupling characteristic of TMD excitons, and to assess the possibility of optical spin control. Another interesting direction is to explore the doping dependence of these emitters in gated heterostructures, and their coupling with a spin-valley–polarized electron gas, which can be optically generated in tungsten-based monolayers \cite{robert-2021} — potentially allowing electrical and/or optical tuning of the emitter charge state and spin properties.

\section{Conclusion}

We have shown that high-temperature annealing accesses a distinct defect–exciton regime in monolayer WS$_2$, enabling the activation of spectrally isolated quantum emitters. This regime is qualitatively different from strain- or disorder-induced localization and points to thermal activation of intrinsic defect states as the dominant mechanism. Our results establish in situ high-temperature annealing as a controllable pathway to access defect-bound excitonic states and single-photon emission in van der Waals materials. Importantly, this approach relies on the ability to suspend layered materials on micromembranes, allowing for real-time monitoring of optical properties while annealing at temperatures that are difficult to achieve in conventional bulk semiconductors. As such, the method is broadly applicable to layered materials, including other transition-metal dichalcogenides and hBN.

\section*{Acknowledgements}
We thank H. Duprez and S. Annaby for their valuable help and technical assistance regarding sample preparation. We also thank M.Srivastava who made the very first annealing attempts during her Bachelor thesis. FC acknowledges support from the European Research Council (ERC Starting Grant OneSPIN, Grant No. 101075855). FC and FS acknowledge the financial support by DEEP2D (ANR-22-CE09-0013). K.W. and T.T. acknowledge support from the JSPS KAKENHI (Grant Numbers 21H05233 and 23H02052) , the CREST (JPMJCR24A5), JST and World Premier International Research Center Initiative (WPI), MEXT, Japan. 

\section*{Data Availability}
The data that support the findings of this study are available in Zenodo at 10.5281/zenodo.17950833.

\section*{Competing Interests}
 The authors declare that
there are no competing interests.

\section*{Supporting information}
The Supplemental Material includes:
\begin{itemize}
  \item Raman spectra as a function of annealing temperature.
  \item Images showing the stability of the membrane's lateral position and of the dissociation of monolayer WS$_2$ above 1000$^\circ$ C.
  \item Analysis of the blackbody radiation of the SiC membrane
  \item Other defect-related emission observed in several samples.
  \item Analysis of the power dependence of the defect emission under pulsed laser excitation.
 \end{itemize}

 \bibliography{acs-template}

@article{mitterreiter-2021,
  title     = {The Role of Chalcogen Vacancies for Atomic Defect Emission in {{MoS2}}},
  author    = {Mitterreiter, Elmar and Schuler, Bruno and Micevic, Ana and {Hernang{\'o}mez-P{\'e}rez}, Daniel and Barthelmi, Katja and Cochrane, Katherine A. and Kiemle, Jonas and Sigger, Florian and Klein, Julian and Wong, Edward and Barnard, Edward S. and Watanabe, Kenji and Taniguchi, Takashi and Lorke, Michael and Jahnke, Frank and Finley, Johnathan J. and Schwartzberg, Adam M. and Qiu, Diana Y. and {Refaely-Abramson}, Sivan and Holleitner, Alexander W. and {Weber-Bargioni}, Alexander and Kastl, Christoph},
  year      = {2021},
  month     = jun,
  journal   = {Nature Communications},
  volume    = {12},
  number    = {1},
  pages     = {3822},
  publisher = {Nature Publishing Group},
  issn      = {2041-1723},
  doi       = {10.1038/s41467-021-24102-y},
  langid    = {english}
}

@article{tonndorf-2015,
  title     = {Single-Photon Emission from Localized Excitons in an Atomically Thin Semiconductor},
  author    = {Tonndorf, Philipp and Schmidt, Robert and Schneider, Robert and Kern, Johannes and Buscema, Michele and Steele, Gary A. and {Castellanos-Gomez}, Andres and van der Zant, Herre S. J. and de Vasconcellos, Steffen Michaelis and Bratschitsch, Rudolf},
  year      = {2015},
  month     = apr,
  journal   = {Optica},
  volume    = {2},
  number    = {4},
  pages     = {347--352},
  publisher = {Optica Publishing Group},
  issn      = {2334-2536},
  doi       = {10.1364/OPTICA.2.000347},
  langid    = {english}
}

@article{ugeda-2014,
  title     = {Giant Bandgap Renormalization and Excitonic Effects in a Monolayer Transition Metal Dichalcogenide Semiconductor},
  author    = {Ugeda, Miguel M. and Bradley, Aaron J. and Shi, Su-Fei and {da Jornada}, Felipe H. and Zhang, Yi and Qiu, Diana Y. and Ruan, Wei and Mo, Sung-Kwan and Hussain, Zahid and Shen, Zhi-Xun and Wang, Feng and Louie, Steven G. and Crommie, Michael F.},
  year      = {2014},
  month     = dec,
  journal   = {Nature Materials},
  volume    = {13},
  number    = {12},
  pages     = {1091--1095},
  publisher = {Nature Publishing Group},
  issn      = {1476-4660},
  doi       = {10.1038/nmat4061},
  langid    = {english}
}

@article{liang-2021,
  title     = {Defect {{Engineering}} of {{Two-Dimensional Transition-Metal Dichalcogenides}}: {{Applications}}, {{Challenges}}, and {{Opportunities}}},
  author    = {Liang, Qijie and Zhang, Qian and Zhao, Xiaoxu and Liu, Meizhuang and Wee, Andrew T. S.},
  year      = {2021},
  month     = feb,
  journal   = {ACS Nano},
  volume    = {15},
  number    = {2},
  pages     = {2165--2181},
  publisher = {American Chemical Society},
  issn      = {1936-0851},
  doi       = {10.1021/acsnano.0c09666}
}

@article{velicky-2017,
  title   = {From Two-Dimensional Materials to Their Heterostructures: {{An}} Electrochemist's Perspective},
  author  = {Velick{\'y}, Mat{\v e}j and Toth, Peter S.},
  year    = {2017},
  month   = sep,
  journal = {Applied Materials Today},
  series  = {{{2D Materials}} in {{Electrochemistry}}},
  volume  = {8},
  pages   = {68--103},
  issn    = {2352-9407},
  doi     = {10.1016/j.apmt.2017.05.003}
}

@article{zinkiewicz-2021b,
  title     = {Excitonic {{Complexes}} in N-{{Doped WS2 Monolayer}}},
  author    = {Zinkiewicz, Ma{\l}gorzata and Wo{\'z}niak, Tomasz and Kazimierczuk, Tomasz and Kapuscinski, Piotr and Oreszczuk, Kacper and Grzeszczyk, Magdalena and Barto{\v s}, Miroslav and Nogajewski, Karol and Watanabe, Kenji and Taniguchi, Takashi and Faugeras, Clement and Kossacki, Piotr and Potemski, Marek and Babi{\'n}ski, Adam and Molas, Maciej R.},
  year      = {2021},
  month     = mar,
  journal   = {Nano Letters},
  volume    = {21},
  number    = {6},
  pages     = {2519--2525},
  publisher = {American Chemical Society},
  issn      = {1530-6984},
  doi       = {10.1021/acs.nanolett.0c05021}
}

@article{nagler-2018a,
  title     = {Zeeman {{Splitting}} and {{Inverted Polarization}} of {{Biexciton Emission}} in {{Monolayer}} \$\{{\textbackslash}mathrm\{\vphantom{\}\}}{{WS}}\vphantom\{\}\vphantom\{\}\_\{2\}\$},
  author    = {Nagler, Philipp and Ballottin, Mariana V. and Mitioglu, Anatolie A. and Durnev, Mikhail V. and Taniguchi, Takashi and Watanabe, Kenji and Chernikov, Alexey and Sch{\"u}ller, Christian and Glazov, Mikhail M. and Christianen, Peter C. M. and Korn, Tobias},
  year      = {2018},
  month     = aug,
  journal   = {Physical Review Letters},
  volume    = {121},
  number    = {5},
  pages     = {057402},
  publisher = {American Physical Society},
  doi       = {10.1103/PhysRevLett.121.057402}
}

@article{goryca-2019,
  title     = {Revealing Exciton Masses and Dielectric Properties of Monolayer Semiconductors with High Magnetic Fields},
  author    = {Goryca, M. and Li, J. and Stier, A. V. and Taniguchi, T. and Watanabe, K. and Courtade, E. and Shree, S. and Robert, C. and Urbaszek, B. and Marie, X. and Crooker, S. A.},
  year      = {2019},
  month     = sep,
  journal   = {Nature Communications},
  volume    = {10},
  number    = {1},
  pages     = {4172},
  publisher = {Nature Publishing Group},
  issn      = {2041-1723},
  doi       = {10.1038/s41467-019-12180-y},
  langid    = {english}
}

@article{srivastava-2015,
  title     = {Optically Active Quantum Dots in Monolayer {{WSe2}}},
  author    = {Srivastava, Ajit and Sidler, Meinrad and Allain, Adrien V. and Lembke, Dominik S. and Kis, Andras and Imamo{\u g}lu, A.},
  year      = {2015},
  month     = jun,
  journal   = {Nature Nanotechnology},
  volume    = {10},
  number    = {6},
  pages     = {491--496},
  publisher = {Nature Publishing Group},
  issn      = {1748-3395},
  doi       = {10.1038/nnano.2015.60},
  langid    = {english}
}

@article{hotger-2023,
  title     = {Spin-Defect Characteristics of Single Sulfur Vacancies in Monolayer {{MoS2}}},
  author    = {H{\"o}tger, A. and Amit, T. and Klein, J. and Barthelmi, K. and Pelini, T. and Delhomme, A. and Rey, S. and Potemski, M. and Faugeras, C. and Cohen, G. and {Hernang{\'o}mez-P{\'e}rez}, D. and Taniguchi, T. and Watanabe, K. and Kastl, C. and Finley, J. J. and {Refaely-Abramson}, S. and Holleitner, A. W. and Stier, A. V.},
  year      = {2023},
  month     = apr,
  journal   = {npj 2D Materials and Applications},
  volume    = {7},
  number    = {1},
  pages     = {30},
  publisher = {Nature Publishing Group},
  issn      = {2397-7132},
  doi       = {10.1038/s41699-023-00392-2},
  langid    = {english}
}

@article{barthelmi-2020,
  title   = {Atomistic Defects as Single-Photon Emitters in Atomically Thin {{MoS2}}},
  author  = {Barthelmi, K. and Klein, J. and H{\"o}tger, A. and Sigl, L. and Sigger, F. and Mitterreiter, E. and Rey, S. and Gyger, S. and Lorke, M. and Florian, M. and Jahnke, F. and Taniguchi, T. and Watanabe, K. and Zwiller, V. and J{\"o}ns, K. D. and Wurstbauer, U. and Kastl, C. and {Weber-Bargioni}, A. and Finley, J. J. and M{\"u}ller, K. and Holleitner, A. W.},
  year    = {2020},
  month   = aug,
  journal = {Applied Physics Letters},
  volume  = {117},
  number  = {7},
  pages   = {070501},
  issn    = {0003-6951},
  doi     = {10.1063/5.0018557}
}

@article{fournier-2021,
  title     = {Position-Controlled Quantum Emitters with Reproducible Emission Wavelength in Hexagonal Boron Nitride},
  author    = {Fournier, Clarisse and Plaud, Alexandre and Roux, S{\'e}bastien and Pierret, Aur{\'e}lie and Rosticher, Michael and Watanabe, Kenji and Taniguchi, Takashi and Buil, St{\'e}phanie and Qu{\'e}lin, Xavier and Barjon, Julien and Hermier, Jean-Pierre and Delteil, Aymeric},
  year      = {2021},
  month     = jun,
  journal   = {Nature Communications},
  volume    = {12},
  number    = {1},
  pages     = {3779},
  publisher = {Nature Publishing Group},
  issn      = {2041-1723},
  doi       = {10.1038/s41467-021-24019-6},
  langid    = {english}
}

@article{micevic-2022,
  title   = {On-Demand Generation of Optically Active Defects in Monolayer {{WS2}} by a Focused Helium Ion Beam},
  author  = {Micevic, A. and Pettinger, N. and H{\"o}tger, A. and Sigl, L. and Florian, M. and Taniguchi, T. and Watanabe, K. and M{\"u}ller, K. and Finley, J. J. and Kastl, C. and Holleitner, A. W.},
  year    = {2022},
  month   = nov,
  journal = {Applied Physics Letters},
  volume  = {121},
  number  = {18},
  pages   = {183101},
  issn    = {0003-6951},
  doi     = {10.1063/5.0118697}
}

@article{fartas-2025,
  title   = {Reproducible Generation of Green-Emitting Color Centers in {{hBN}} Using Oxygen Annealing},
  author  = {Fartas, Helmi and Hassani, Sa{\"i}d and Hermier, Jean-Pierre and Lai, Ngoc Diep and Buil, St{\'e}phanie and Delteil, Aymeric},
  year    = {2025},
  month   = jul,
  journal = {Applied Physics Letters},
  volume  = {127},
  number  = {1},
  pages   = {014001},
  issn    = {0003-6951},
  doi     = {10.1063/5.0261073}
}

@article{castellanos-gomez-2014,
  title     = {Deterministic Transfer of Two-Dimensional Materials by All-Dry Viscoelastic Stamping},
  author    = {{Castellanos-Gomez}, Andres and Buscema, Michele and Molenaar, Rianda and Singh, Vibhor and Janssen, Laurens and {van der Zant}, Herre S J and Steele, Gary A},
  year      = {2014},
  month     = apr,
  journal   = {2D Materials},
  volume    = {1},
  number    = {1},
  pages     = {011002},
  publisher = {IOP Publishing},
  issn      = {2053-1583},
  doi       = {10.1088/2053-1583/1/1/011002},
  langid    = {english}
}

@article{cadiz-2017,
  title     = {Excitonic {{Linewidth Approaching}} the {{Homogeneous Limit}} in \$\{{\textbackslash}mathrm\{\vphantom{\}\}}{{MoS}}\vphantom\{\}\vphantom\{\}\_\{2\}\$-{{Based}} van Der {{Waals Heterostructures}}},
  author    = {Cadiz, F. and Courtade, E. and Robert, C. and Wang, G. and Shen, Y. and Cai, H. and Taniguchi, T. and Watanabe, K. and Carrere, H. and Lagarde, D. and Manca, M. and Amand, T. and Renucci, P. and Tongay, S. and Marie, X. and Urbaszek, B.},
  year      = {2017},
  month     = may,
  journal   = {Physical Review X},
  volume    = {7},
  number    = {2},
  pages     = {021026},
  publisher = {American Physical Society},
  doi       = {10.1103/PhysRevX.7.021026}
}

@article{paur-2019,
  title     = {Electroluminescence from Multi-Particle Exciton Complexes in Transition Metal Dichalcogenide Semiconductors},
  author    = {Paur, Matthias and {Molina-Mendoza}, Aday J. and Bratschitsch, Rudolf and Watanabe, Kenji and Taniguchi, Takashi and Mueller, Thomas},
  year      = {2019},
  month     = apr,
  journal   = {Nature Communications},
  volume    = {10},
  number    = {1},
  pages     = {1709},
  publisher = {Nature Publishing Group},
  issn      = {2041-1723},
  doi       = {10.1038/s41467-019-09781-y},
  langid    = {english}
}

@article{passler-1997,
  title   = {{Basic Model Relations for Temperature Dependencies of Fundamental Energy Gaps in Semiconductors}},
  author  = {P{\"a}ssler, R.},
  year    = {1997},
  journal = {physica status solidi (b)},
  volume  = {200},
  number  = {1},
  pages   = {155--172},
  issn    = {1521-3951},
  doi     = {10.1002/1521-3951(199703)200:1<155::AID-PSSB155>3.0.CO;2-3},
  langid  = {french}
}

@article{guo-2015,
  title   = {Chalcogen Vacancies in Monolayer Transition Metal Dichalcogenides and {{Fermi}} Level Pinning at Contacts},
  author  = {Guo, Y. and Liu, D. and Robertson, J.},
  year    = {2015},
  month   = apr,
  journal = {Applied Physics Letters},
  volume  = {106},
  number  = {17},
  pages   = {173106},
  issn    = {0003-6951},
  doi     = {10.1063/1.4919524}
}

@book{brainard-1968,
  title       = {The {{Thermal Stability}} and {{Friction}} of the {{Disulfides}}, {{Diselenides}}, and {{Ditellurides}} of {{Molybdenum}} and {{Tungsten}} in {{Vacuum}} (10{\=9} to 10{\=6} {{Torr}})},
  author      = {Brainard, William A.},
  year        = {1968},
  publisher   = {{National Aeronautics and Space Administration}},
  googlebooks = {hdfWZ0J0o60C},
  langid      = {english}
}

@article{cadiz-2016a,
  title     = {Ultra-Low Power Threshold for Laser Induced Changes in Optical Properties of {{2D}} Molybdenum Dichalcogenides},
  author    = {Cadiz, Fabian and Robert, Cedric and Wang, Gang and Kong, Wilson and Fan, Xi and Blei, Mark and Lagarde, Delphine and Gay, Maxime and Manca, Marco and Taniguchi, Takashi and Watanabe, Kenji and Amand, Thierry and Marie, Xavier and Renucci, Pierre and Tongay, Sefaattin and Urbaszek, Bernhard},
  year      = {2016},
  month     = oct,
  journal   = {2D Materials},
  volume    = {3},
  number    = {4},
  pages     = {045008},
  publisher = {IOP Publishing},
  issn      = {2053-1583},
  doi       = {10.1088/2053-1583/3/4/045008},
  langid    = {english}
}

@misc{huberich-2025,
  title        = {Atomically-Resolved Exciton Emission from Single Defects in {{MoS}}\$\_2\$},
  author       = {Huberich, Lysander and Ammerman, Eve and Yu, Gu and Ren, Yining and Papadopoulos, Sotirios and Dong, Chengye and Robinson, Joshua A. and Watanabe, Kenji and Taniguchi, Takashi and Gr{\"o}ning, Oliver and Novotny, Lukas and Li, Tingxin and Wang, Shiyong and Schuler, Bruno},
  year         = {2025},
  month        = oct,
  journal      = {arXiv.org},
  howpublished = {https://arxiv.org/abs/2510.15676v1},
  langid       = {english}
}

@article{robert-2021,
  title     = {Spin/Valley Pumping of Resident Electrons in {{WSe2}} and {{WS2}} Monolayers},
  author    = {Robert, Cedric and Park, Sangjun and Cadiz, Fabian and Lombez, Laurent and Ren, Lei and Tornatzky, Hans and Rowe, Alistair and Paget, Daniel and Sirotti, Fausto and Yang, Min and Van Tuan, Dinh and Taniguchi, Takashi and Urbaszek, Bernhard and Watanabe, Kenji and Amand, Thierry and Dery, Hanan and Marie, Xavier},
  year      = {2021},
  month     = sep,
  journal   = {Nature Communications},
  volume    = {12},
  number    = {1},
  pages     = {5455},
  publisher = {Nature Publishing Group},
  issn      = {2041-1723},
  doi       = {10.1038/s41467-021-25747-5},
  langid    = {english}
}

@article{carbone-2025,
  title   = {Creation and Microscopic Origins of Single-Photon Emitters in Transition-Metal Dichalcogenides and Hexagonal Boron Nitride},
  author  = {Carbone, Amedeo and {Bendixen-Fernex de Mongex}, Diane-Pernille and Krasheninnikov, Arkady V. and Wubs, Martijn and Huck, Alexander and Hansen, Thomas W. and Holleitner, Alexander W. and Stenger, Nicolas and Kastl, Christoph},
  year    = {2025},
  month   = sep,
  journal = {Applied Physics Reviews},
  volume  = {12},
  number  = {3},
  pages   = {031333},
  issn    = {1931-9401},
  doi     = {10.1063/5.0278132}
}

@article{sun-2024,
  title     = {Unveiling Sulfur Vacancy Pairs as Bright and Stable Color Centers in Monolayer {{WS2}}},
  author    = {Sun, Huacong and Yang, Qing and Wang, Jianlin and Ding, Mingchao and Cheng, Mouyang and Liao, Lei and Cai, Chen and Chen, Zitao and Huang, Xudan and Wang, Zibing and Xu, Zhi and Wang, Wenlong and Liu, Kaihui and Liu, Lei and Bai, Xuedong and Chen, Ji and Meng, Sheng and Wang, Lifen},
  year      = {2024},
  month     = nov,
  journal   = {Nature Communications},
  volume    = {15},
  number    = {1},
  pages     = {9476},
  publisher = {Nature Publishing Group},
  issn      = {2041-1723},
  doi       = {10.1038/s41467-024-53880-4},
  langid    = {english}
}

@article{vancso-2016,
  title     = {The Intrinsic Defect Structure of Exfoliated {{MoS2}} Single Layers Revealed by {{Scanning Tunneling Microscopy}}},
  author    = {Vancs{\'o}, P{\'e}ter and Magda, G{\'a}bor Zsolt and Pet{\H o}, J{\'a}nos and Noh, Ji-Young and Kim, Yong-Sung and Hwang, Chanyong and Bir{\'o}, L{\'a}szl{\'o} P. and Tapaszt{\'o}, Levente},
  year      = {2016},
  month     = jul,
  journal   = {Scientific Reports},
  volume    = {6},
  number    = {1},
  pages     = {29726},
  publisher = {Nature Publishing Group},
  issn      = {2045-2322},
  doi       = {10.1038/srep29726},
  langid    = {english}
}

\end{document}